


\magnification=1200
\baselineskip=20pt
\rightline{NHCU-HEP-94-23}
\centerline{\bf Nonlocal Matrix Generalizations of N=2 Super Virasoro Algebra}
\vskip 1.5cm
\centerline{Wen-Jui Huang}
\vskip 1cm
\centerline{Department of Physics}
\centerline{National Tsing Hua University}
\centerline{Hsinchu, Taiwan, Republic of China}
\vskip 1.5cm
\centerline{\bf \underbar{Abstract}}
\vskip 1cm

We study the generalization of the second Gelfand-Dickey bracket to the
superdifferential
operators with matrix-valued coefficients. The associated matrix
Miura transformation is derived. Using this bracket we work out a nonlocal and
nonlinear N=2 superalgebra which contains the N=2 super Virasoro algebra as a
subalgebra. The bosonic limit of this superalgebra is considered. We show that
when the spin-1 fields in this bosonic algebra are set to zero the resulting
Dirac bracket gives precisely the recently derived $V_{2,2}$ algebra.

\vfil\eject

\noindent{\bf 1. Introduction}

The connections between Hamiltonian structures of integrable systems and
classical extended conformal algebras are now well understood [1-4]. The
extensions of
 conformal algebras by higher spin fields and their supersymmetric versions are
known to given by the Gelfand-Dickey brackets associated with
appropriated (super) pseudodifferential operators [5-10].

Recently, in the study of simplest 1+1 dimensional non-abelian Toda field
theory, an interesting nonlocal extended conformal algebra called V algebra  is
discovered [11].
This algebra consists of a Virasoro generator $t$ and two additional spin-2
fields $v_{\pm}$:
$$\eqalign{ \{t(x), t(y) \} &= [- {1 \over 2} \partial_x^3 + 2 t(x) \partial_x
 + t'(x)] \delta (x-y) \cr
\{v_{\pm} (x), t(y) \} &= [ 2 v_{\pm}(x) \partial_x + v'_{\pm}(x) \delta (x-y)
\cr
\{ v_{\pm}(x), v_{\pm}(y) \} &= 2 \epsilon (x-y) v_{\pm}(x) v_{\pm} (y) \cr
\{ v_{\pm}(x), v_{\mp}(y) \} &= -2 \epsilon (x-y) v_{\pm}(x) v_{\mp}(y) \cr
      &\qquad + [- {1 \over 2} \partial_x^3 + 2 t(x) \partial_x + t'(x)]
\delta (x-y) \cr} \eqno(1)$$
where $\epsilon(x-y) = \partial_x^{-1} \delta(x-y)$ (This differs from the
definition used in Ref.[11] by a factor of 2.)
It is realized that this algebra is actually given by the second
Gelfand-Dickey bracket associated with the second order differential operator
with $2 \times 2$ matrix-valued coefficients:
$$ L = \partial^2 - U \eqno(2)$$
where
$$ U = \pmatrix{ t  &-\sqrt{2} v_+ \cr
	       - \sqrt{2} v_-  &t\cr} \eqno(3)$$
In this correspondence, it is understood that the nonlocal terms arise as
a consequnce of the noncommutative character of matrix multiplication [11].
This discovery leads to the generalization of the second Gelfand-Dickey
to the differential operators with matrix-valued coefficients. The resulting
algebras are nonlocal matrix generalization of $W$ algebras [12].

The purpose of this paper is to study the supersymmetric version of the above
generalization. More pecisely, we consider the second Gelfand-Dickey bracket
associated superdifferential operators with matrix-valued coifficients. The
corresponding matrix Muira transformation is presented and used to prove the
Jocobi identity for the second Gelfand-Dickey bracket. We work out
explicitly the
bracket associated with the third order superdifferential operators with
$ 2 \times 2 $ matrix-valued coefficients. The resulting superalgebra, which
contains the classical N=2 Super Virasoro algebra as a subalgebra, is a
nonlocal matrix generalization of N=2 super Virasoro algebra.
We then consider further the bosonic limit of this superalgebra.
In particular, we show explicitly that the $V_{2,2}$ algebra obtained in
Ref.[12]
arises from this bosonic algebra as the Dirac bracket when the spin-1 fields
are all set to zero.
\vskip 1cm

\noindent{\bf 2. Second Gelfand-Dickey Bracket and Its Miura Transformation}

The needed ingredients	to generalize the second Gelfand-Dickey bracket to
the case of the superdifferential operators with matrix-valued coefficients are
already contained in Refs.[7,12].
Hence, the generalization is quite
straightforward and we shall be very brief.

We consider the superdifferential operators defined on $(1|1)$ superspace with
coordinates $X=(x,\theta)$:
$$ L = D^m + U_1 D^{m-1} + \dots + U_m \eqno(4)$$
where $D=\partial_{\theta} + \theta \partial_x$ and $U_i$'s are $n \times n$
matrices whose entries are N=1 superfields. It is always assumed that the
operators are homogeneous under $Z_2$ grading; that is, $|U_i|=i(mod 2)$.
For later uses we need to recall some basic definitions. First, we denote
the Berezin integral
by $\int_B=\int dx d\theta$ with the convention $\int d\theta \theta =1$.
The super Leibnitz rule for the covariant derivative $D$ is
$$ D^k \Phi = \sum_{i=0}^{\infty} \left[ \matrix{k \cr k-i} \right]
(-1)^{|\Phi|(k-i)} \Phi^{[i]} D^{k-i}  \eqno(5)$$
where k is an integer and $\Phi^{[i]}=(D^i \Phi)$ [{\it Note:} We also use
the notations: $\Phi'=\Phi^{[1]}$ and $\Phi''=\Phi^{[2]}$, etc. If $f(x)$ is
an ordinary function, then $f'(x)$ means $\partial_x f(x)$],
and the superbinomial
coefficients $\left[ \matrix{k \cr k-i} \right]$ are defined by
$$ \left[ \matrix{k \cr k-i} \right] = \left\{ \eqalign{ &0, \qquad for \quad
i<0 \qquad or \quad (k,k-i) \equiv (0,1) \quad (mod \quad 2) \cr
& \pmatrix{ \left[ {k \over 2} \right] \cr \left[ {k-i \over 2} \right]},
\qquad otherwise, \cr} \right\} \eqno(6)$$
where $\pmatrix{p \cr q}$ is the ordinary binomial coefficient. Given a
superpseudodifferential operator $P=\sum p_i D^i$, where $p_i$'s are $n
\times n$ matrices, we define its super-residue as $sres P = p_{-1}$. It can
be shown easily
$$ \int_B tr \quad sres[P,Q] = 0 \eqno(7)$$
where $tr$ is the ordinary trace of matrix and $[P,Q] \equiv PQ- (-1)^{|P||Q|}
QP$ is the supercommutator.

Given a functional $F[U] = \int_B f(U)$, where $f(U)$ is a $Z_2$-homogeneous
differential polynomial of $U_i$'s, we define its matrix-valued gradient $X_F$
by
$$ X_F = \sum_{i=1}^m (-1)^i D^{-m+i-1} {{\delta f} \over {\delta U_k}}
\eqno(8)$$
where the matrix ${{\delta f} \over {\delta U_k}}$ is defined by
$$ \left( {{\delta f} \over {\delta U_k}} \right)_{ij} = \sum_{l=0}^{\infty}
(-1)^{|U_k|l +l(l+1)/2} D^l {{\partial f} \over {\partial (U_k)^{[l]}_{ji}}}
 \eqno(9)$$
With these definitions we now define the second Gelfand-Dickey bracket as
$$ \{F, G \} = (-1)^{|F|+|G|+m} \int_B tr \quad sres [ L (X_F L)_+ X_G- (L
X_F)_+ L X_G]  \eqno(10)$$
where $(\quad)_+$ denotes the differetial part of a superpseudodifferential
operator. The antisupersymmetry of the bracket is obvious by the virtue of the
construction. The super Jacobi identity will follow from the Miura
transformation which we shall introduce now.

Let us consider m $n \times n$ matrices $\Phi_i$'s whose entries are
grassmanian odd N=1 superfields. We introduce a Poisson bracket defined by
$$\eqalign{ \{ F, G \}^* &\equiv \sum_{i=1}^m (-1)^{m+i+1} \int_B tr \left\{
\left({{\delta f} \over {\delta \Phi_i}} \right)' {{\delta g} \over {\delta
\Phi_i}} - \Phi_i \left[ {{\delta f} \over {\delta \Phi_i}}, {{\delta g}
\over {\delta \Phi_i}} \right] \right\} \cr
&= \sum_{i=1}^m (-1)^{m+i+1} \int_B tr \left\{ \left[ \nabla_i, {{\delta f}
\over {\delta \Phi_i}} \right] {{\delta g} \over {\delta \Phi_i}} \right\} \cr}
\eqno(11)$$
where $F[\Phi]= \int_B f(\Phi)$, $G[\Phi]= \int_B g(\Phi)$ and $\nabla_i \equiv
D - \Phi_i$. One can show easily that the bracket (11) is indeed
antisupersymmetric and satisfies the super Jacobi identity. As in the scalar
case the two brackets (10) and (11) are actually equivalent once we make
the following identification
$$ \eqalign{ L &= D^m + U_1 D^{m-1} + \dots + U_m \cr
	       &= (D-\Phi_1)(D-\Phi_2) \dots (D-\Phi_m) \cr} \eqno(12)$$
Eq.(12) is the desired matrix Miura transformation. With the use of (7) the
proof for this equivalenc is almost a repetition of that for the scalar case.
We
shall not spell it out here.

Since our main purpose is to get a supersymmetric version of V algebra,
we have to consider the constraint $U_1=0$. Let us first discuss the constraint
in the context of the bracket (11). From the matrix Miura transformation (12)
we have
$$ U_1 = \sum_{i=1}^m (-1)^i \Phi_i \eqno(13)$$
To see whether this constraint is of second class or not we consider two
functionals $F$ and $G$ which depend on $\Phi_i$'s only through $U_1$; that is,
$f$ and $g$ are functions of $U_1$ only. By (13) we get ${{\delta f} \over
{\delta \Phi_i}} = (-1)^i {{\delta f} \over {\delta U_1}}$ and a similar
expression for $g$. Hence,
$$\{F, G \}^*|_{U_1=0} = \sum_{i=1 }^m (-1)^{m+i+1} \int_B tr
\left\{\left(
{{\delta f} \over {\delta U_1}} \right)' {{\delta g} \over {\delta U_1}}
\right\}_{U_1=0} \eqno(14)$$
Clearly, the constraint is of second class only when
m is odd. In this case the constraint imposed on ${{\delta f} \over {\delta
\Phi_i}}$'s is
$$ \sum_{i=1}^m \left[ \nabla_i, {{\delta f} \over {\delta \Phi_i}} \right]
_{U_1=0} = 0 \eqno(15)$$
Following the procedure used in Refs.[7,12] we can trasnslate (15) into a
constraint on $X_F$:
$$ sres [ L, X_F ]|_{U_1=0} = 0 \eqno(16)$$
which is more useful for computing brackets.

{}From now on we shall concern only with superdifferential operators of odd
order
$$L= D^{2m+1} + U_2 D^{2m-1} + \dots + U_{2m+1} \eqno(17)$$
We note that if $F$ and $G$ depend linearly on $U_i$'s through the traces
then $X_F$ and $X_G$ are both proportional to the $n \times n$ idnetity matrix
and hence the computations of the brackets are similar to the scalar case.
In particular, if we define $J=tr(U_2)$ and $T=tr(U_3)-{1 \over 2} tr(U'_2)$
we find
$$\eqalign{ \{T(X), T(Y) \} &= [{1 \over 4} nm(m+1) D^5 + {3 \over 2} T(X)D^2
+ {1 \over 2} T'(X) + T''(X)] \delta(X-Y) \cr
\{J(X), T(Y) \} &= [J(X)D^2 - {1 \over 2}J'(X) D + J''(X)] \delta(X-Y) \cr
\{J(X), J(Y) \} &= -[nm(m+1)D^3 +2 T(X)] \delta(X-Y) \cr} \eqno(18)$$
where $\delta(X-Y) \equiv \delta(x-y)(\theta_X-\theta_Y)$. Eq.(18) is the
classical N=2 super Virasoro algebra. We shall see later that the condition
(16) leads to the presence of nonlocal terms in the resulting  algebra.
In other words, the second
Gelfand-Dickey bracket associated with the operator (17) gives a nonlocal
matrix generalization of classical N=2 W-superalgebra.
\vskip 1cm

\noindent{\bf 3. A Nonlocal Extension of Classical N=2 Super Virasoro Algebra}

Now we shall work out in details the second Gelfand-Dickey bracket associated
with the following operator
$$ L = D^3 + U_2 D + U_3 \eqno(19)$$
To this purpose it is convinient to evaluate first
$$ J(X) \equiv L(XL)_+ - (LX)_+ L \eqno(20)$$
where $X=D^{-3} X_1 + D^{-2} X_2 + D^{-1} X_3$. Staightforward algebras give
$$ J(X) = J_1 D^2 + J_2 D + J_3$$
where
$$\eqalign{ J_1 &= (-1)^{|X|} \left\{ -X'_1 + X''_2 + X'''_3 + [U_2, X_2]
-[U_3, X_3] -(X_3 U_2)' \right\} \cr
J_2 &= X''_1 + X_2''' +[U_2, X_1] + U_2 X_2' -(X_2 U_2)' + U_3 X_2 + U_3 X_3'
+ (-1)^{|X|} X_2 U_3 \cr
J_3 &= (-1)^{|X|} \big\{-X'''_1 + X_3^{[5]} + U_2 (X_3'''-X_1') -[U_3,X_1] +
U_3 X''_3   - (-1)^{|X|} (X_3 U_3)'' \cr
&\qquad- (X_3 U_2)'''  - (-1)^{|X|}
(X_2 U_3)' -U_2(X_3 U_2)' - U_3 X_3 U_2 - (-1)^{|X|} U_2 X_3 U_3 \big\} \cr}
\eqno(21)$$
Due to the constraint $U_1=0$ the coefficient of $D^2$, $J_1$, must vanish.
This requirement, of course, is equivalent to the condition (16). We thus
have $X_1 = X_2' + X_3'' -X_3 U_2 + (D^{-1} [U_2,X_2]) - (D^{-1} [U_3, X_3])$.
Putting this into the expressions of $J_2$ and $J_3$ we have
$$J_2 = J_{22} + J_{23} \qquad\qquad J_3 = J_{32} + J_{33}$$
where
$$\eqalign{ J_{22} &= 2 X_2''' - (-1)^{|X|} X_2 U_2 + [U_2', X_2] + 3 [U_2,
X_2']
+ U_3 X_2 \cr
 &\qquad + (-1)^{|X|} X_2 U_3 + [U_2,(D^{-1} [U_2, X_2])] \cr
J_{23} &= X_3^{[4]} + (X_3 U_2)'' + U_3 X_3' + [U_2, X_3''] - [U_3, X_3]' \cr
 &\qquad -[U_2, X_3 U_2] - [U_2, (D^{-1} [U_3, X_3]) ] \cr
J_{32} &= (-1)^{|X|} \big\{ - X_2^{[4]} - U_2 X_2'' - (-1)^{|X|} (X_2 U_3)' -
[U_2, X_2]'' - [U_3, X_2'] \cr
 &\qquad - U_2 [U_2, X_2] - [U_3, (D^{-1}[U_2, X_2])] \big\} \cr
J_{33} &= (-1)^{|X|} \big\{ U_3 X_3'' - (-1)^{|X|} (X_3 U_3)'' + [U_3'', X_3]
+[U_2 U_3, X_3] \cr
 &\qquad  + [U_3, (D^{-1}[U_3, X_3])] \big\} \cr } \eqno(22)$$

Using (8), (10) and (22) we can easily write down the desired brackets:
$$\eqalign{ &\{tr[fU_2(X)], tr[gU_2(Y)] \} \cr
&\quad= tr \big\{ -2 fg D_X^3 - 2 gf
 [U_3(x) - {1 \over 2} U_2'(X)] \big\} \delta(X-Y) + tr \big\{ [f,g] \big[
3 U_2(X) D_x  \cr
&\qquad - U_3(X) +2 U_2'(X) \big] + [U_2,f] D^{-1}_X [U_2,g] \big\}
\delta(X-Y) \cr}$$
$$\eqalign{ &\{tr[fU_2(X)], tr[gU_3(Y)] \} \cr
&\quad= tr \big\{ -fg D^4_X + gf \big[ U_2(X) D^2_X
-U_3(X) D_X + U''_2(X) \big] \big\} \delta(X-Y) \cr
&\qquad + tr \big\{ [f,g] \big[ 2 U_2(X) D^2_X +U_3(X) D_X
 - U_3'(X) + U_2''(X) \big] -U_2(X) [U_2(X), f] g \cr
&\qquad - [U_2(X),f] D^{-1}_X [U_2(X),g] \big\} \delta(X-Y) \cr
&\{tr[fU_3(X)], tr[gU_3(Y)] \} \cr
 &\quad =  tr \big\{ 2 gfU_3(X) D^2_X + gf U''_3(X) \big\} \delta(X-Y) +
 tr \big\{ [f,g] \big( U_3(X) D^2_X \cr
&\quad - U_2(X) U_3(X) \big) - [U_3(X),f] D^{-1}_X [U_3(X),g] \big\}
\delta(X-Y) \cr} \eqno(23)$$
Here $f$ and $g$ are two constant real-valued matrices. We note that if
either
of $f$ and $g$ is proportional to the identity matrix then those terms
involving at least a commutator would vanish and the resulting brackets
resemble   those in the scalar case. For example, if we define
$$ W_2(X) \equiv U_2(X)  \qquad \qquad W_3(X) \equiv U_3(X) - {1 \over 2}
U'_2(X) \eqno(24)$$
then, similar  to the scalar case, $tr(fW_2)$ and $tr(fW_3)$ form a N=2
supermultiplet, provided $f$ is a
traceless matrix. In other words, we have the following brackets:
$$\eqalign{ \{ tr[fW_2(X)], J(Y) \} &= -2 tr[fW_3(X)] \delta(X-Y) \cr
\{ tr[fW_3(X)], J(Y) \} &= tr \big\{ f[-W_2(X) D^2_X + {1 \over 2} W'_2(X)
D_X - {1 \over 2} W''_2(X)] \big\} \delta(X-Y) \cr
\{ tr[fW_2(X)], T(Y) \} &= tr \big\{ f[W_2(X) D^2_X - {1 \over 2 } W'_2(X)
D_X + W''_2(X)] \big\} \delta(X-Y) \cr
\{ tr[fW_3(X)], T(Y) \} &= tr \big\{ f[{3 \over 2} W_3(X) D^2_X + {1 \over 2}
W'_3(X) D_X + W''(X)] \big\} \delta(X-Y) \cr} \eqno(25)$$
We can also write down the other brackets. However, to be more specific,
we shall focus on the case of $2 \times 2$ matrix. We parametrize the matrices
$W_2$
and $W_3$ as follows
$$ W_2 = \pmatrix{ {1 \over 2} J + {1 \over 2} J_3  & J_+ \cr
 J_-  &{1 \over 2} J - {1 \over 2} J_3 \cr}
\qquad W_3 = \pmatrix{ {1 \over 2} T + {1 \over 2} W_3  &W_+ \cr
W_-  &{1 \over 2} T - {1 \over 2} W_3 \cr} \eqno(26)$$
Using (24) and the notation $\in (X-Y) \equiv D^{-1}_X \delta(X-Y)$ we obtain
$$\eqalign{ \{ J_{\pm}(X), J_{\pm}(Y) \} &= 2 \in (X-Y) J_{\pm}(X) J_{\pm}(Y)
\cr
\{ J_{\pm}(X), J_{\mp}(Y) \} &= - \in (X-Y) [2 J_{\pm}(X) J_{\mp}(Y) +
J_3(X) J_3(Y)] \cr
&\quad -[2 D^3_X + T(X)] \delta(X-Y) \mp [ 3 J_3(X) D_X + {3 \over 2} J'_3(X)]
\delta(X-Y) \cr
\{ J_3(X), J_{\pm}(Y) \} &= 2 \in (X-Y) J_{\pm}(X) J_3(Y) \mp [6 J_{\pm}(X)
D_X + 3 J'_{\pm}(X)] \delta (X-Y) \cr
\{ J_3(X), J_3(Y) \} &= -4 \in (X-Y) [J_+(X) J_-(Y) + J_-(X) J_+(Y)] \cr
&\quad -2[2 D^3_X + T(X)] \delta (X-Y) \cr}$$
$$\eqalign{ \{J_{\pm}(X), W_{\pm}(Y) \} &= 2 \in (X-Y) J_{\pm}(X) W_{\pm}(Y)
\cr
\{ J_{\pm} (X), W_{\mp}(Y) \} &= - \in (X-Y) [ 2 J_{\pm}(X) W_{\mp} (Y) +
J_3(X) W_3(Y)] \cr
&\quad + {1 \over 2} [ J(X) D^2_X
-{1 \over 2} J'(X) D_X + J''(X) ] \delta (X-Y) \cr
 &\quad \mp [ {3 \over 2}
W_3(X) D_X - W'_3(X) - { 1 \over 2} J(X) J_3(X) ] \delta(X-Y) \cr
\{ J_{\pm}(X), W_3(Y) \} &= 2 \in (X-Y) J_3(X) W_{\pm} (Y) \cr
&\qquad \pm [3 W_{\pm}(X) D_X - 2 W'_{\pm}(X) - J(X) J_{\pm}(X)] \delta(X-Y)
\cr
\{ J_3(X), W_{\pm}(Y) \} &= 2 \in (X-Y) J_{\pm}(X) W_3(Y) \cr
&\qquad \mp [3 W_{\pm}(X) D_X - 2 W'_{\pm}(X) - J(X) J_{\pm}(X)] \delta(X-Y)
\cr
\{J_3(X), W_3(Y) \} &= -4 \in (X-Y) [J_+(X) W_-(Y) + J_-(X) W_+(Y)]
+  \cr
&\quad + [ J(X) D^2_X - {1 \over 2} J'(X) D_X + J''(X) ] \delta(X-Y) \cr}
\eqno(27)$$
$$\eqalign{ \{ W_{\pm}(X), W_{\pm}(Y) \} &= 2 \in (X-Y) W_{\pm}(X) W_{\pm}(Y)
- {1 \over 2} J_{\pm}(X) D_X J_{\pm} (X) \delta (X-Y) \cr
\{ W_{\pm}(X), W_{\mp}(Y) \} &= - \in (X-Y) [2 W_{\pm}(X) W_{\mp}(Y) + W_3(X)
W_3(Y)]  \cr
&\quad + {1 \over 2} [ D^5_X + {3 \over 2} T(X) D^2_X + {1 \over 2} T'(X) D_X
+ T''(X) ] \delta(X-Y) \cr
&\qquad [ {1 \over 2} J_{\pm}(X) D_X J_{\mp}(X) + {1 \over 4} J_3(X) D_X J_3(X)
] \delta (X-Y) \cr
&\quad \pm {1 \over 2} [J(X) W_3(X) + J_3(X) T(X)] \delta(X-Y) \cr
&\qquad \pm [{3 \over 4} J_3(X) D^3_X + {3 \over 8} J'_3(X) D^2_X + {3 \over 8}
J''_3(X) D_X] \delta(X-Y) \cr}$$
$$\eqalign{ \{ W_3(X), W_{\pm}(Y) \} &= 2 \in (X-Y) W_{\pm} (X) W_3(Y) - {1
\over 2}
J_{\pm}(X) D_X J_3(X) \delta(X-Y) \cr
&\qquad \pm [J(X) W_{\pm}(X) + J_{\pm}(X) T(X) ] \delta(X-Y) \cr
&\qquad \pm [ {3 \over 2} J_{\pm}(X) D^3_X + {3 \over 4} J'_{\pm}(X) D^2_X
+ {3 \over 4} J''_{\pm}(X) D_X ] \delta(X-Y) \cr
\{W_3(X), W_3(Y) \} &= - 4 \in (X-Y) [ W_+(X) W_-(Y) + W_-(X) W_+(Y)] \cr
&\quad + [D^5_X + {3 \over 2} T(X) D^2_X + {1 \over 2} T'(X) D_X + T''(X) ]
\delta(X-Y) \cr
&\qquad + [J_+(X) D_X J_-(X) + J_-(X) D_X J_+(X) ] \delta(X-Y) \cr}$$
We like to remark here that the superalgebra given by (27) is invariant under
the transformations:
$$\eqalign{ J_{\pm} \longrightarrow J_{\mp}  \qquad & J_3 \longrightarrow - J_3
\cr
 W_{\pm} \longrightarrow W_{\mp} \qquad & W_3 \longrightarrow - W_3 \cr}
\eqno(28)$$
The transformations (28) is due to the invariance of the second Gelfand-Dickey
bracket under the constant similarity transformation:
$$ L \longrightarrow S^{-1} L S  \qquad  X_{F(G)} \longrightarrow S^{-1}
X_{F(G)} S \eqno(29)$$
Here $S$ is a constant nonsingular matrix. Indeed, if we take $S = \pmatrix{
0 \quad 1 \cr 1 \quad 0 \cr}$, then the first of (29) gives (28).

The other remark is that the Dirac bracket arising from setting $J$ and $J_a
(a= 3, \pm)$ to zero would involve correction terms proportional to
$[2 D^3 + T]^{-1}$. This makes these brackets to have infinite
number of nonlocal terms. Hence, this reduction is not well defined.
Therefore how  to get a nonlocal matrix generalization of N=1 Super Virasoro
algebra remains an interesting question.
\vskip 1cm

\noindent{\bf 4. $V_{2,2}$ algebra As Dirac Bracket}

In the end of the last section we see that eliminating the spin-1 superfields
does not give a well defined N=1 superalgebra. However,
eliminating the fermionic fields in a superalgebra is quite straightforward. To
get the bosonic limit
of the N=2 superalgebra defined by (18), (25), (26)  and (27) we write the
superfields in component form:
$$\eqalign{ T(X) &= \phi(x) - \theta_X t(x) \qquad  J(X) = j(x) + \theta_X
\varphi(x) \cr
W_a(X) &= \phi_a(x) + \theta_X v_a(x)  \qquad J_a(X) = j_a(x) + \theta_X
\varphi_a(x) \cr} \eqno(30)$$
Setting $\phi$ and $\varphi$ to be zero in (18) (with n=2 and m=1) we have
$$\eqalign{ \{ t(x), t(y) \} &= [ - \partial_x^3 + 2 t(x) \partial_x +
t'(x) ] \delta (x-y) \cr
\{ j(x), t(y) \} &= [ j(x) \partial_x + j'(x) ] \delta(x-y) \cr
\{ j(x), j(y) \} &= -4 \partial_x \delta(x-y) \cr} \eqno(31)$$
which is, as expected, the statement that $t$ is the Virasoro generator and $j$
is a spin-1 field. While (25) gives  ( for $a= \pm, 3$)

$$\eqalign{ \{ j_a(x), j(y) \} &= 0   \cr
       \{ v_a(x), j(y) \} &= [ -j_a(x) \partial_x ] \delta(x-y) \cr
\{ j_a(x),  t(y) \} &= [ j_a(x) \partial_x + j'_a(x) ] \delta(x-y) \cr
\{ v_a(x), t(y) \} &= [2 v_a(x) \partial_x + v'_a(x) ] \delta(x-y) \cr}
\eqno(32)$$
Finally, the bosonic limit of (27) are
$$\eqalign{ \{ j_{\pm}(x), j_{\pm}(y) \} &= 2 \epsilon(x-y) j_{\pm}(x) j_{\pm}
(y) \cr
\{ j_{\pm}(x), j_{\mp}(y) \} &= - \epsilon(x-y) [ 2 j_{\pm}(x) j_{\mp}(y) +
j_3(x) j_3(y) ] \delta(x-y) \cr
&\qquad - 2 \delta'(x-y) \mp 3 j_3(x) \delta(x-y) \cr
\{ j_3(x), j_{\pm}(y) \} &= 2 \epsilon(x-y) j_{\pm}(x) j_3(y) \mp 6 j_{\pm}(x)
\delta(x-y) \cr
\{ j_3(x), j_3(y) \} &= -4 \epsilon(x-y) [j_+(x) j_-(y) + j_-(x) j_+(y) ]
-4 \delta'(x-y) \cr
\{ j_{\pm}(x), v_{\pm}(y) \} &= 2 \epsilon(x-y) j_{\pm}(x) v_{\pm}
(y) \cr
\{ j_{\pm}(x), v_{\mp}(y) \} &= - \epsilon(x-y) [2 j_{\pm}(x) v_{\mp}(y) +
j_3(x) v_3(y) ] - {1 \over 2} [j(x) \partial_x + j'(x)] \delta(x-y) \cr
&\qquad \mp [v_3(x) + {1 \over 2} j(x) j_3(x) ] \delta(x-y) \cr
\{ j_{\pm}(x), v_3(y) \} &= 2 \epsilon(x-y) j_3(x) v_{\pm}(y) \pm [2 v_{\pm}(x)
+ j(x) j_{\pm}(x) ] \delta(x-y) \cr
\{ j_3(x), v_{\pm}(y) \} &= 2 \epsilon(x-y) j_{\pm}(x) v_3(y) \mp [2 v_{\pm}(x)
+ j(x) j_{\pm}(x) ] \delta(x-y) \cr
\{j_3(x), v_3(y) \} &= -4 \epsilon(x-y) [j_+(x) v_-(y) + j_-(x) v_+(y)]
  - [j(x) \partial_x + j'(x) ] \delta(x-y) \cr}\eqno(33)$$

$$\eqalign{ \{   v_{\pm}(x), v_{\pm}(y) \} &= 2 \epsilon(x-y) v_{\pm}(x)
v_{\pm}(y)
 + {1 \over 2} j_{\pm}(x) \partial_x j_{\pm}(x) \delta(x-y) \cr
\{ v_{\pm}(x), v_{\mp}(y) \} &= - \epsilon(x-y) [2 v_{\pm(x)} v_{\mp}(y) +
v_3(x) v_3(y) ] + {1 \over 2} [ - \partial_x^3 + 2 t(x) \partial_x \cr
&\quad + t'(x)] \delta(x-y) -[{1 \over 2} j_{\pm}(x)  \partial_x j_{\mp}(x) +
{1 \over 4} j_3(x) \partial_x j_3(x) ] \delta(x-y) \cr
&\quad \mp [{3 \over 4} j_3(x) \partial_x^2 + {3 \over 4} j'_3(x) \partial_x
+ {1 \over 2} j(x) v_3(x) + {1 \over 2} j_3(x) t(x)] \delta(x-y) \cr
\{ v_3(x), v_{\pm}(y) \} &= 2 \epsilon(x-y) v_{\pm}(x) v_3(y) + {1 \over 2}
j_{\pm}(x) \partial_x j_3(x) \delta(x-y) \cr
&\quad \mp [ {3 \over 2} j_{\pm}(x) \partial_x^2 + {3 \over 2} j'_{\pm}(x)
\partial_x
+ j(x) v_{\pm}(x) + j_{\pm}(x) t(x) ] \delta(x-y) \cr
\{v_3(x), v_3(y) \} &= -4 \epsilon(x-y) [v_+(x) v_-(y) + v_-(x) v_+(y)] + [
- \partial_x^3 \cr
&\qquad + 2 t(x) \partial_x + t'(x) ] \delta(x-y) - [j_+(x) \partial_x j_-(x)
+ j_-(x) \partial_x j_+(x) ] \delta(x-y) \cr}$$
This algebra contains, besides the Virasoro generator $t$, three spin-2 fields,
$v_a(a= \pm,3)$ and four spin-1 fields, $j$ and $j_a(a =\pm, 3)$.  The
four spin-1 fields form a closed subalgebra. In fact, the local parts of the
brackets $\{ j_a(x), j_b(x) \}$ define nothing but the ordinary $sl(2,R)$ Kac-
Moody algebra. Hence, the first four brackets of (33) give a example of
nonlocal generalization of Kac-Moody algebra.

Since the $V_{2,2}$ algebra consists of a Virasoro generator and three spin-2
fields, it is natural to expect that it can be obtained from the above algebra
by setting the four spin-1 fields to be zero. We now show that this
expectation is actually true. From (31) we see that $\{ j_a(x), j(y) \} =0 $
and $\{v_a(x), j(y) \}|_{j_a=0} = 0$. A little thinking tells us that
to take care of the constraint $j=0$ we simply drop
$j$-dependent terms in (33). Assuming that this has been done
we note further that $\{ v_a(x), j_b(y)
\}|_{j_{\pm},j_3 =0} $ are not all zero. Some corrections of the
brackets must be taken. Since the $3 \times 3$ matrix
$\{j_a(x), j_b(y) \}|_{j_{\pm},j_3 =0 }$ is formally invertible, we can simply
use the formula of Dirac bracket:
$$\eqalign{ &\{v_a(x), v_b(y) \}^{Dirac} = \{v_a(x), v_b(y) \}|_{j_{\pm},j_3
=0} \cr
&\qquad  - \sum_{c,d = \pm,3} \int \int dw dz \{v_a(x),j_c(w) \}|_{j_{\pm},j_3
=0}
M_{c,d}(w,z) \{j_d(z), v_b(y) \}|_{j_{\pm},j_3 =0 } \cr} \eqno(34)$$
where $M_{c,d}(w,z)$ is the inverse of the matrix $\left( \{j_c(w), j_d(z) \}
|_{j_{\pm},j_3 =0} \right)$. The second piece is the correction to the original
bracket
due to the constraints $j_{\pm},j_3 = 0$. In operator form it reads
$$\eqalign{ &\qquad { - \pmatrix{ 0  &-v_3  &2v_+ \cr
	       v_3  &0  &-2v_- \cr
	       -2v_+  &2v_-  &0 \cr}
\pmatrix{ 0  &-{1 \over 2} \partial^{-1}  &0 \cr
	 -{1 \over 2} \partial^{-1}  &0  &0 \cr
	 0  &0  &-{1 \over 4} \partial^{-1} \cr}
\pmatrix { 0  &-v_3  &2v_+ \cr
	   v_3  &0  &-2v_- \cr
	  -2v_+  &2v_-  &0 \cr} } \cr
&= - \pmatrix {v_+ \partial^{-1} v_+ \quad &-{1 \over 2} v_3 \partial^{-1} v_3
 -v_+ \partial^{-1} v_- \quad &v_3 \partial^{-1} v_+ \cr
-{1 \over 2} v_3 \partial^{-1} v_3 - v_- \partial^{-1} v_+ \quad &v_-
\partial^{-1} v_- \quad &v_3 \partial^{-1} v_- \cr
v_+ \partial^{-1} v_3 \quad &v_- \partial^{-1} v_3 -2 v_- \partial^{-1} v_+
\quad &- 2 v_+ \partial^{-1} v_- \cr} \cr}\eqno(35)$$
Each correction term is purely nonlocal and, in fact, equals to
negative of one half of the nonlocal term in the corresponding uncorrected
bracket. Therefore the resulting Dirac brackets read
$$ \eqalign{ \{ v_{\pm}(x), v_{\pm}(y) \}^{Dirac} &= \epsilon(x-y) v_{\pm}(x)
v_{\pm}(y) \cr
\{ v_{\pm}(x), v_{\mp}(y) \}^{Dirac} &= - \epsilon(x-y) [ v_{\pm}(x) v_{\mp}(y)
+ {1 \over 2} v_3(x) v_3(y) ] \cr
&+ {1 \over 2} [- \partial_x^3 + 2 t(x) \partial_x + t'(x)  ] \delta(x-y) \cr
\{ v_3(x) , v_{\pm}(y) \}^{Dirac} &= \epsilon(x-y) v_{\pm}(x) v_3(y) \cr
\{ v_3(x), v_3(y) \}^{Dirac} &= -2 \epsilon(x-y) [v_+(x) v_-(y) + v_-(x)
v_+(y)]
\cr
&+ [ - \partial_x^3 + 2 t(x) \partial_x + t'(x) ] \delta(x-y) \cr}\eqno(36)$$
which, together with the first bracket in (31) and the last bracket in  (32),
define the $V_{2,2}$ algebra. More precisely, it is the second Gelfand-Dickey
bracket associated with $L= \partial^2 - U$, where $U$ is parametrized by
$$ U = \pmatrix { {1 \over 2} t + {1 \over 2 } v_3  &v_+ \cr
       v_-  &{1 \over 2} t - {1 \over 2} v_3 \cr} \eqno(37)$$
Of course, $V_{2,2}$ algebra can be regarded as the Dirac bracket arising from
the local algebra defined by the second Gelfand-Dickey bracket associated with
$L = \partial^2 + V \partial -	U$ by setting the $2 \times 2$ matrix $V$ to
zero.
What we have shown here is that it can also arise from a nonlocal algebra.
\vskip 1cm

 This work was supported by the National Science Council of Republic of
China under Grant number NSC-83-0208-M-007-008.

\vskip 1cm

\noindent{\bf References:}
\item{[1]} A. Das, {\it Integrable Models} (World Scientific, 1988); L. Dickey,
{\it Soliton Equations and Hamiltonian Systems } (World
Scientific, 1991).
\item{[2]} I.M. Gelfand and L.A. Dickey, Funct. Anal. Appl. {\bf 11}, 93
(1977);
 M. Adler, Invent. Math. {\bf 50}, 219 (1979); B.A. Kuperschmidt and G. Wilson,
 Invent. Math. {\bf 62}, 403 (1981).
\item{[3]} V.G. Drinfeld and V.V. Sokolov, J. Sov. Math. {\bf30}, 1975 (1984).
\item{[4]} P. Bouwknegt and K. Schoutens, Phys. Rep. {\bf 223}, 183 (1993) and
references therein.
\item{[5]} J.-L. Gervais, Phys. Lett. {\bf B160}, 277 (1985); P. Mathieu, Phys.
 Lett. {\bf B208}, 101 (1988); I. Bakas, Comm. Math. Phys. {\bf 123}, 627
(1989).
\item{[6]} P. Di Francesco, C. Itzykson and J.-B. Zuber, Comm. Math. Phys.
{\bf 140}, 543 (1991).
\item{[7]} J.M. Figeroa-O'Farrill, J. Mas and E. Ramos, Phys. Lett. {\bf
299}, 41 (1993).
\item{[8]} J.M. Figeroa-O'Farrill and E. Ramos, Phys. Lett. {\bf
B262}, 265 (1991); Nucl. Phys. {\bf B368}, 361 (1992).
\item{[9]} T. Inami and H. Kanno, Nucl. Phys. {\bf B359}, 201 (1991); J. Phys.
{\bf A25}, 3729 (1992).
\item{[10]} F. Gieres and S. Theisen, J. Math. Phys. {\bf 34}, 5964 (1993);
W.-J. Huang, J. Math. Phys. {\bf 35}, 2570 (1994).
\item{[11]} A. Bilal, {\it Non Abelian Toda Theory: A Completely Integrable
Model for Strings on A Black Hole Background}, preprint PUPT-1434 (hep-th/
9312108); {\it Multi-Component KdV Hierarchy, V Algebra and Non Abelian Toda
Theory}, preprint PUPT-1446 (hep-th/9401167).
\item{[12]} A. Bilal, {\it Non-Local Matrix Generalizations of W-Algebras},
preprint PUPT-1452 (hep-th/9403197).

\end

\end